# NCI Workshop on Artificial Intelligence in Radiation Oncology: Training the Next Generation

John Kang, MD, PhD. Dept. of Radiation Oncology, University of Rochester Medical Center, NY

Reid F. Thompson, MD PhD, Dept. of Radiation Medicine, Oregon Health & Science University, Portland OR; Staff Physician, VA Portland Healthcare System

Sanjay Aneja, MD. Dept of Therapeutic Radiology, Yale University, New Haven, NY

Constance Lehman, MD, PhD Dept. of Radiology, Harvard Medical School, Mass General Hospital, Boston MA

Andrew Trister, MD, PhD, Bill & Melinda Gates Foundation, Seattle, WA

James Zou, PhD. Dept. of Biomedical Data Science, Stanford University. Chan-Zuckerberg Biohub. San Francisco, CA

Ceferino Obcemea, PhD (co-chair). Radiation Research Program, National Cancer Institute, Bethesda, MD

Issam El Naqa, PhD (co-chair), Department of Radiation Oncology, University of Michigan, Ann Arbor, MI

Corresponding author:
    John Kang, MD, PhD. 601 Elmwood Ave, Rochester, NY 14642
    johnkan1@alumni.cmu.edu, 585-275-2171

## Table of Contents

# Abstract

Artificial intelligence (AI) is about to touch every aspect of radiotherapy from consultation, treatment planning, quality assurance, therapy delivery, to outcomes modeling. There is an urgent need to train radiation oncologists and medical physicists in data science to help shepherd AI solutions into clinical practice. Poorly trained personnel may do more harm than good when attempting to apply rapidly developing and complex technologies. As the amount of AI research expands in our field, the radiation oncology community needs to discuss how to educate future generations in this area. The National Cancer Institute (NCI) Workshop on AI in Radiation Oncology (Shady Grove, MD, April 4-5, 2019) was the first (https://dctd.cancer.gov/NewsEvents/20190523_ai_in_radiation_oncology.htm) of two data science workshops in radiation oncology hosted by the NCI in 2019. During this workshop, the Training and Education Working Group was formed by volunteers among the invited attendees. Its members represent radiation oncology, medical physics, radiology, computer science, industry, and the NCI.

In this perspective article written by members of the Training and Education Working Group, we provide and discuss Action Points relevant for future trainees interested in radiation oncology AI: (1) creating AI awareness and responsible conduct; (2) implementing a practical didactic curriculum; (3) creating a publicly available database of training resources; and (4) accelerate learning and funding opportunities. Together, these Action Points can facilitate the translation of AI into clinical practice.

# Introduction

Artificial intelligence (AI) is a longstanding field of study that has attempted to emulate and augment human intelligence. In the last several years, AI has been reinvigorated by advances in computer technology and machine learning (ML) algorithms, which aim to teach computers to learn patterns and rules by using previous examples. ML builds on experiences from computer science, statistics, neuroscience, and control theory, among many other disciplines. ML has benefited from recent availability of large datasets and developments in computers' hardware and software for solving large-scale optimization problems. Most notably, deep learning (DL) techniques have demonstrated significant successes in computer vision and language processing. These advances are most visible in consumer quality-of-life improvements such as self-driving cars and voice-activated virtual assistants. The umbrella term "informatics" includes practical applications of any of the above areas of study; for example, bioinformatics for biology and clinical informatics (or biomedical informatics) for clinical practice. The term "data science" refers to the general study of data analysis, which has recently focused on ML methods. A schematic of the relationships between common terminologies is shown in **Figure 1**.

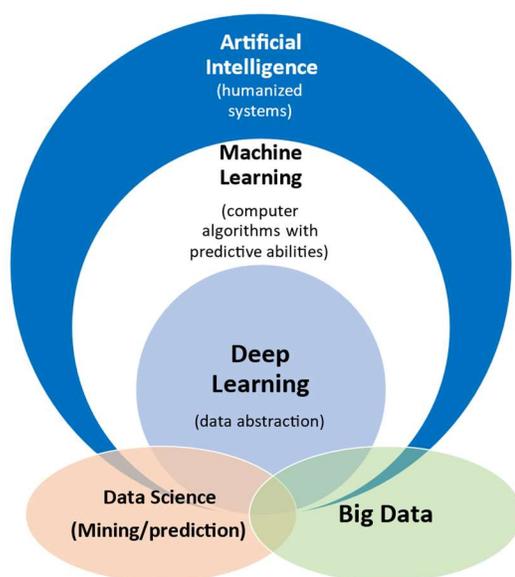

**Figure 1:** Schematic of how artificial intelligence (AI), machine learning (ML), and deep learning (DL) relate to each other. Closely associated application areas such as "data analytics" and "big data" exist both within and outside of these realms. Salient examples include: AI includes expert systems using rules (if-then statements) and statistical ML; ML includes support vector machines and neural networks; DL includes deep neural networks and convolutional neural networks; big data can be described as data having volume, velocity, variety, veracity/variability, and value (https://www.ibm.com/blogs/watson-health/the-5-vs-of-big-data/); data analytics refers to the process of making meaningful predictions and models, as exemplified by the work of several authors referenced in this paper[24,25,33,41,43,47].

Many fields such as finance, manufacturing, and advertising have already incorporated AI into their workflows to improve efficiency and perform supra-human tasks. While AI has been adopted more slowly in the clinic due to multiple competing factors—including a lack of training—the perception and engagement of AI in medicine has been improving. The American Medical Association (AMA) adopted a policy in June 2019 to integrate training in AI augmentation [1]. The National Institutes of Health (NIH) Big Data to Knowledge (BD2K) Initiative has established several Centers of Excellence in Data Science, and is focused on enhancing nationwide training infrastructure in biomedical data science as well as data sharing [2]. Radiation oncology holds significant promise for AI-powered tasks—described in several perspectives and reviews [3–8]—not just for optimizing workflows or diagnosis, but also more rewarding tasks such as prognostic prediction and personalized treatment recommendations.

AI applications in radiation oncology span the domains of both medical physicists and radiation oncologists. Some applications, such as auto-segmentation and automated treatment planning, will be human-verifiable; in other words, a human can check the work of a computer prior to deployment. Other applications—survival prognostication, decision support, and genomics-based treatment planning are not human-verifiable at an individual scale and will thus require careful model development and validation. As applied research in these applications grows in radiation oncology, a commensurate growth in education is necessary to be able to build and validate trustworthy AI models that can be applied to the clinic.

Separate surveys of trainees in radiation oncology and radiology reveal that the majority are interested in additional training in the AI or informatics [9,10].

In the radiation oncology survey, the American Society of Radiation Oncology (ASTRO) queried chairs and trainees in 2017 to assess their perception of training and research opportunities in genomics, bioinformatics, and immunology.[10] Among the three areas, bioinformatics received the most enthusiasm: 76% believed that bioinformatics training would "definitely or probably" advance their career. 67% expressed interest in a formal bioinformatics training course and 88% of chairs reported they would "probably or definitely" send faculty or trainees to such a course, reflecting an unmet need in training opportunities. Though the ASTRO survey did not specifically ask about AI/ML, we believe the high interest in bioinformatics accurately reflects interest in quantitative analysis in line with AI/ML methods[11]. In recognition of the need for radiation oncologists with specialized informatics training, the NCI, Oregon Health & Science University, and MD Anderson Cancer Center have each created training programs specific for radiation oncology fellows/residents aimed for careers as medical director in informatics and/or formal board certification in clinical informatics [12].

The radiology survey supports a sentiment towards AI that is similar to that in radiation oncology [13]. A single institution survey of a radiology department revealed concerns about job security but also enthusiasm to learn about AI/ML [9]. This survey showed that 97% of trainees (residents and fellows) were planning to learn AI/ML as relevant to their job (vs. 77% of attending radiologists). In fact, 74% of trainees (vs. 60% of attendings) were willing to help create or train an ML algorithm to do some of the tasks as a radiologist. National radiology societies have been responsive to these sentiments. The American College of Radiology (ACR) Data Science Institute (https://www.acrdsi.org/) recently launched the ACR AI-LAB™ to allow radiologists to create, validate and use models for their specific local clinical needs. The Radiological Society of North America (RSNA) and the Society for Imaging Informatics in Medicine (SIIM) co-sponsor the National Imaging Informatics Course, held twice a year (in its 3rd year); the majority of residency programs have participated (https://sites.google.com/view/imaging-informatics-course/). The RSNA annual meeting hosts several AI refresher courses and coding challenges in the new "AI Pavilion" with residents encouraged to participate. The SIIM Resident, Fellow, Doctoral Candidate, Student (RFDS) society hosts monthly journal clubs and promotes mentoring opportunities (https://siim.org/page/rfds_community_inter).

In this perspective article written by the Training and Education Working Group of the NCI Workshop on AI in Radiation Oncology (Shady Grove, MD, April 4-5, 2019) [14], we propose an overall action plan for radiotherapy-specific AI training that is comprised of the action points outlined in **Table 1**. We cover each action point (AP) in detail in this article.

**Table 1: Main summary of recommendations from Action Points**

| |
|---|
| **AP1: Create awareness and responsible conduct of AI**<br>● Teach the importance of consideration of ethics, disparities, bias, and fairness in AI |
| **AP2: Implement practical didactic curriculum**<br>● Curriculum should address the needs of medical physicists, radiation oncologists, and data scientists and their respective roles in the process.<br>● Mandate incorporation of AI, Big Data, data privacy, and data science training into residency curricula<br>● Reiterate the importance of promoting a data sharing culture through education and exposure |
| **AP3: Create publicly accessible resources**<br>● Create centralized, public repository of resources (seminal white papers, webinars, video-lectures, hackathons, datasets, code, etc.)<br>● Develop and disseminate radiation-specific training tools to all institutions to democratize access and standardize training quality<br>● Leverage trainee crowdsourcing efforts to annotate datasets for research and use annotated datasets for education towards model and AI skills development<br>● Incorporate open source challenges for both technology development and educational purposes |
| **AP4: Accelerate learning and funding opportunities**<br>● Incorporate workshop model used by AACR, ESTRO and others for rapid learning opportunities<br>● Leverage industry support and interest via AI fellowships and tutorial workshops<br>● Identify applicable research grants specifically for trainees/new investigators |

**Abbreviations**: AACR: American Association for Cancer Research; AI: artificial intelligence; ESTRO: European Society for Radiotherapy and Oncology

## AP1: Create awareness and responsible conduct of AI

In the last decade or so, we have seen several examples of ethical concerns and biases magnified by AI. When there are biases in the training data (e.g., certain populations or scenarios are over represented), then an algorithm that models correlations could propagate or even amplify these biases, leading to undesirable outcomes in deployment [15]. This is particularly problematic as AI is sometimes viewed as being "objective" without consideration for the data generation process, which is often unknown.

The European Union (EU) has recently released a seven-point action plan towards so-called "trustworthy" AI. This plan focuses on the ethical aspects of AI and includes: human agency and oversight; robustness and safety;

privacy and data governance; transparency; diversity, non-discrimination and fairness; societal and environmental well-being; and accountability (https://ec.europa.eu/futurium/en/ai-alliance-consultation/guidelines). Similarly, the Food and Drug Administration (FDA) has taken similar steps towards regulation of AI applications in medicine [16]. A key component of improving awareness is to be transparent and clearly document where and when an AI algorithm is used in any part of the clinical workflow. And in cases where AI is applied, researchers and physicians should also clarify whether the AI is an ML system⎯which are the more recent type of AI trained on large data and tend to be less interpretable⎯or an older rule-based system. ML and rule-based AI have different behaviors. For example, neural networks—a type of ML system—are vulnerable to adversarial attacks [17,18].

## AP2: Implement practical didactic curriculum

There are currently no educational guidelines for AI training in radiation oncology or medical physics residents. Serendipitously, there is an active discussion within the field about revising the radiation oncology resident training curriculum. While in depth discussion of all the factors at play is beyond the scope of discussion here, we refer readers to a pair of editorials by Amdur and Lee [19] and Wallner et. al [20]. In July 2020, the Accreditation Council for Graduate Medical Education made several changes to the radiation oncology residency curriculum.[21] The revisions are notable for mandating education in several new areas, including clinical informatics. We are pleased that the ACGME has the foresight to update training curriculum to include informatics and hope that this paper can serve to provide high-level guidance.

In **Action Point 2**, we propose a high-level overview of a curriculum draft for trainees in medical physics and radiation oncology to adequately grasp the basic principles of AI. These principles are generalizable to medicine as a whole and have particular significance for interventional and informatics-heavy specialties such as radiation oncology.

1. Responsible conduct of AI, bias, and disparities
2. Methodology: data science basics

3. Interpreting data and models
4. Practical experience and applications
5. Data sharing: logistics and culture

## 2.1 Responsible conduct of AI, bias, and disparities

There is increasing concern that AI models influenced by bias will further perpetuate healthcare disparities for patients. The underlying reason behind why bias is retained in AI models is often related to training data which fails to represent the entire population equally. Because AI algorithms do not have a concept of "fairness", surveillance of inherent bias with AI is typically left to those who designed the system. As noted by the EU/FDA in **Action Point 1**, proper application of AI should aim to enhance positive social change and enhance sustainability and ecological responsibility. Particularly in medicine, rules and regulations should be put in place to ensure responsibility and accountability of AI systems, their users and their appropriate utilization. In the computer science and ML communities, there has been increasing efforts to improve the teaching of ethics and human-centered AI in coursework (https://stanfordcs181.github.io/)[22]. A complementary area of work is to develop methods to audit AI systems in order to identify potential systematic or cultural biases. Trainees must develop an appreciation for these critical complexities and potential limitations of AI.

## 2.2 Methodology: data science basics

Data features, structures, and algorithms form the foundation of AI applications. Unfortunately, quantitative analysis and critical data appraisal are not universally emphasized in medical or post-graduate education, particularly for physicians. As many ML techniques become published in general medical or oncology journals, it is incumbent upon editors and readers alike to have some basic facility with the techniques. Building a working knowledge of basic statistical concepts such as hypothesis testing, confidence intervals, and basic performance metrics will need to be introduced before more data structures and model-agnostic techniques such as data cleaning, cross validation, model fitting, bias-variance trade off, and advanced performance metrics, such as the widely-used but poorly-understood receiver operating curve[23]. To de-mystify many of these topics, there are existing high-quality online courses made broadly available, which will be further discussed in **Action Point 2** and **Action Point 3**.

## 2.3 Interpreting data and models

For proper clinical application of AI tools, physicians should be able to assess the validity of the data and the model-generation process. So-called "black box" models have such internal complexity that they are conceptualized as inputs mapped to outputs without any intent to understand how the mapping occurs. Several ML methods, including deep learning (DL) and most ensemble methods, fall into this categorization. While black box AI models can have excellent performance during training and internal validation, they often encounter problems generalizing when widely deployed. Understanding *why* a problem occurred can be difficult with "black box" models and is currently a very active area of AI research [18,24].

One way to demonstrate data and model interpretability is through "use cases." In medical research, there are well-known examples of the potential dangers of black box models related to confounding [25]. Fortunately, researchers were able to catch these issues before deploying their models, which may not always be the case for complex datasets with nonobvious confounders.

There is an ongoing discussion on the necessity of AI interpretability by the FDA [16,26] and the informatics community [27]. All authors would agree that elevating the knowledge base of clinicians and physicists will certainly enable more innovation regardless of final regulatory plans.

## 2.4 Practical experience and applications

For trainees interested in applying data science to clinical practice, these opportunities should be encouraged and promoted.

While medical physics and radiation oncology AI curricula could have significant overlap, there will necessarily be focuses on separate domains. In medical physics, instruction may cover methods for auto-segmentation, automated/adaptive treatment planning, and quality assurance. Radiation oncology trainees may be more interested in prognostic predictions and clinical decision support. In the future, as AI takes more of an augmented intelligence role, there should be instruction for physicians for how to decide whether to accept, interrogate or reject recommendations. For example, physicians may need to determine whether there is sufficient rationale to accept an automatically generated plan or treatment recommendation using clinical and dosimetric information.

Several radiation oncology departments have AI/ML researchers who could contribute to a training curriculum. These courses should be jointly taught to both physicists and physicians. We anticipate that common courses and collaboration between trainees in medical physics and radiation oncology will improve translation of AI methods into the clinic. Given that medical physicists already have quantitative training in methods with significant overlap with ML, we anticipate close collaboration between physicists and physicians. Indeed, this is the current status quo in most radiation oncology departments performing AI/ML research.

For departments without access to sufficient resources, online education using so-called MOOCs ("massive open online courses;" a misnomer as they are not necessarily massive or open) and workshop models (see **Action Point 4**) may be more educational to trainees than co-opting faculty without training in AI/ML.

For advanced practitioners, we will discuss data science hackathons and crowdsourcing in **Action Point 3**.

## 2.5 Data sharing: logistics and culture

One of the key aspects of creating robust predictive models is being able to show generalization to novel datasets through a process called external validation, which requires institutions to share data among themselves. The data sharing culture in medicine has been historically tribalistic but has gradually become more collaborative. This dynamic was well exemplified by the backlash to an infamous 2016 editorial (co-authored by then-editor-in-chief of the New England Journal of Medicine) that was viewed as anti-data sharing [28,29]. Unlike in academic medicine, academic AI researchers have a strong open-access culture where pre-print archiving of publications is the norm and datasets are simultaneously published with papers to invite validation. Notably, patients are generally supportive of the sharing of their data and would likely embrace scientific reuse of their data to improve the lives of future patients [30], though we recognize that there are many regulatory limitations to widespread data sharing of this sort. Finding a path for controlled data sharing amongst trusted parties, or more broadly with de-identification schemes could be an important first step in improving the accuracy of AI algorithms.

In this curriculum, we hope to emphasize the efforts in medicine and oncology to promote data sharing (**Table 2**).

**Table 2: Examples of data sharing initiatives in oncology**

| Entity | Est. | Area |
|---|---|---|
| The Cancer Genome Atlas (TCGA) | 2005 | Tumor genomics |
| ACR Imaging Network/TRIAD | 2009 | Clinical trial protocols, datasets, cloud-based data transfer |
| Radiogenomics Consortium | 2009 | Radiotherapy genomics/genetics |
| The Cancer Imaging Archive (TCIA) | 2010 | DICOM, radiomics. Select datasets with genomics, histopathology. |
| ASCO CancerLinQ | 2014 | Quality improvement. Plan for decision support. |
| Project DataSphere | 2014 | Phase III cancer clinical trial patient-level data |
| ACR Data Science Institute | 2017 | Use cases in for development of medical imaging AI |
| NCI Office of Data Sharing | 2018 | Advocacy, establishing standards, defining incentives |

**Abbreviations**. ACR: American College of Radiology; DICOM: Digital Imaging and Communications in Medicine; AI: artificial intelligence; ASCO: American Society of Clinical Oncology; Est.: established year; NCI: National Cancer Institute; TRIAD: Transfer of Images and Data

The NCI is keen on improving data sharing protocols and resources. In 2018, the NCI Office of Data Sharing (ODS) was created and the NIH has recently asked for open comments on a draft policy for data management and sharing [31]. The Cancer Imaging Archive (TCIA) is funded by NCI and allows the sharing of anonymized imaging datasets as well as corresponding clinical and genomic data. A radiology initiative includes the ACR Imaging Network (ACRIN) for clinical trial protocol and dataset sharing through TRIAD (Transfer of Images and Data). The American Society of Clinical Oncology (ASCO) provides another strong example of centralized data sharing in the CancerLinQ project (https://cancerlinq.org/). Radiation oncology currently lacks a specialty-specific centralized platform for data request and sharing. A notable effort to move towards this goal is by American Association of Physicists in Medicine (AAPM) Task Group 263 to standardize nomenclature using structured ontology for data pooling [32].

One approach to overcome data transfer medicolegal/PHI issues is through distributed or federated learning. In this approach, analysis is performed locally and models are transferred (e.g., feature weights) instead of data; this decentralized approach has shown equivalent performance to that using central pooling of data [33,34]. Such innovative approaches for anonymization could be part of a training curriculum to help overcome barriers to data sharing.

Recent years have seen signs of a cultural shift in radiation oncology towards open collaboration and data sharing, along with formalization of key principles in data sharing, namely that data should be FAIR: findable, accessible, interoperable, and reusable. These FAIR guiding principles for scientific data management and stewardship [35] are of utmost importance, and should be discussed with and endorsed for all trainees. In line with FAIR, several radiation oncology academic centers and cooperative groups have contributed datasets to the TCIA [36–38]. Open access journals with a focus on radiation oncology include BMC Radiation Oncology, the Frontiers in Oncology section on radiation oncology, and Advances in Radiation Oncology, which was launched by ASTRO in 2015.

Several coordinating efforts present opportunities to pool ideas and data to promote collaborating, increase power for discovery, and avoid redundancy. Within imaging, these efforts include the aforementioned ACR Data Science Institute for AI in medical imaging, which aims to identify clinically-impactful use cases in radiation oncology, such as auto-segmentation and MRI-derived synthetic CT scans [39]. Within genomics, the Radiogenomics Consortium is a transatlantic cooperative effort pooling American and European cohorts to find genomic markers for toxicity to radiotherapy [40]. Several groups within the consortium are interested in creating ML models to predict toxicity response in radiotherapy [41–44].

Through the proposed curriculum draft of **Action Point 2**, we hope to build a core of trainees for the next generation who can understand and apply data science fundamentals while also understanding ethical considerations and data sharing principles.

# AP3: Create publicly accessible resources

Directly building off the curriculum discussed in **Action Point 2**, the third Action Point relates to the creation of a centralized, publicly accessible repository of resources to guide trainees. These resources could include seminal white papers, video lectures, code samples, and contacts for potential collaborations. To reach the widest potential audience, we favor storage at open access websites such as GitHub and Youtube, for instance. As discussed in **Action Point 2**, a formal curriculum can be facilitated and standardized using MOOCs, which would consist of video lectures and interactive coding exercises. MOOCs have become very influential in online education as they can be tailored for various experience levels and are self-paced. Due to economies of scale, they can be widely disseminated for reasonable costs. For example, the MOOCs on Coursera run on the higher end of cost and charges around $40/month for classes that last around a month with around 10-12 hours of coursework a week. MOOCs could be adopted from existing courses or centrally created in collaboration with organizations such as the Association of Residents in Radiation Oncology (ARRO) Education Committee and Radiation Oncology Education Collaborative Study Group (ROECSG).

As there is more interest, trainees will likely want to be involved in practical research projects. Given that AI expertise is not evenly distributed, both intra- and inter-institutional collaborations can be fostered. In this respect, trainees can provide a valuable service by annotating data for research. At the same time, they would also benefit from the service of others by receiving annotated data for model and skills development. A model for this can be seen in eContour (https://www.econtour.org), a free web-based contouring atlas. In a randomized trial, eContour improved nasopharynx contours and anatomy knowledge compared to traditional resources [45]. A next phase of eContour may involve enabling user-generated contours and segmentations for technology development and research, with a prototype initially pilot-tested at the American College of Radiation Oncology Annual Meeting in 2017 (http://www.econtour.org/acro). Researchers are invited to use the platform to collect contours from large numbers of users from diverse practice locations, though must provide funding to support website programming and content administration. For residents, funding opportunities are available through professional organizations, as discussed in **Action Point 4**.

Another promising venue for trainees interested in skills development is through public competitions (**Table 3**).

**Table 3: Examples of data science competitions in oncology and medicine**

| Platform | Year | Prediction Goal |
|---|---|---|
| MICCAI Brain Tumor Segmentation (BraTS) benchmark | 2012-2020 | Segment heterogeneous brain tumors (gliomas) |
| Prostate Cancer DREAM Challenge | 2015 | Predict overall survival and docetaxel discontinuation in mCRPC |
| Kaggle Data Science Bowl | 2016 | Predict heart ejection fraction |
| MICCAI radiomics challenges (2) | 2016 | (1) HPV (2) local control in oropharyngeal cancer |
| Kaggle Data Science Bowl | 2017 | Detect lung cancer via National Lung Screening Trial DICOMs |
| TopCoder Lung Cancer Challenge | 2017 | Segment lung cancer |
| Kaggle Data Science Bowl | 2018 | Detect cellular nuclei |

**Abbreviations**. DICOM: Digital Imaging and Communications in Medicine; DREAM: Dialogue for Reverse Engineering Assessments and Methods; HPV: human papillomavirus; MICCAI: Medical Image Computing and Computer Assisted Intervention Society; mCRPC: metastatic castrate resistant prostate cancer

Past challenges in radiation oncology have leveraged collaborations between academic centers, international societies such as MICCAI (The Medical Image Computing and Computer Assisted Intervention Society) and commercial sites such as TopCoder.com (Wipro, Bengalaru, India) and Kaggle.com (Google, San Francisco, CA). These public crowd-sourcing challenges have included two radiomics challenges (to predict human papillomavirus status or local control) in oropharyngeal cancer after radiotherapy [46] and to predict lung tumor segmentation [47]. Within the medical computer vision domain, MICCAI hosts several challenges every year to help validate and benchmark image processing algorithms (http://www.miccai.org/events/challenges); regular challenges have included melanoma diagnosis, brain tumor segmentation, and prostate cancer Gleason grading. In the data science competition space as a whole, there has been enthusiasm for healthcare challenges, with the last three Kaggle Data Science Bowls (https://datasciencebowl.com/) focused on heart ejection fraction determination (2016), lung cancer screening (2017), and cellular nuclei detection (2018).

# AP4: Accelerate learning and funding opportunities

Developing and maintaining resources described in **Action Point 3** will require accelerated learning of particularly motivated trainees who will need institutional infrastructure and funding mechanisms to be successful.

While MOOCs provide consistency and quality of education, for accelerated training, the radiation oncology community could adopt the intensive weeklong workshop model that is widely used by oncology organizations [11]. Examples include separate workshops on clinical trial development by ASCO/AACR (https://vailworkshop.org/) and ECCO-AACR-EORTC-ESMO (https://www.ecco-org.eu/Events/MCCR-Workshop/EDITION-20) as well as the AACR Molecular Biology in Clinical Oncology Workshop which focuses on molecular biology methods and grantsmanship (https://www.aacr.org/meeting/mbco-2020/). These intensive week-long workshops are run by established faculty and are aimed at the research career development of senior trainees and junior faculty with a strong focus on mentorship; this workshop model could be adopted in radiation oncology to mentor trainees in data science. The aforementioned ASTRO 2017 survey[10] suggests this workshop model could be well received with 88% of radiation oncology chairs "probably or definitely" be willing to send faculty or trainees to such a course to learn bioinformatics.

**Table 4: Examples of data science workshops in radiation oncology**

| Workshop | Date | Host(s) |
|---|---|---|
| AAPM Practical Big Data Workshop | May 19-20, 2017 | University of Michigan |
| AAPM Practical Big Data Workshop | May 31-June 2, 2018 | University of Michigan |
| EORTC State of Science meeting | September 26-27, 2018 | EORTC |
| AAPM Practical Big Data Workshop | June 6-8, 2019 | University of Michigan |
| NCI Workshop on AI in Radiation Oncology | April 4-5, 2019 | NCI Radiation Research Program |
| NCI State of Data Science in Radiation Oncology | July 25, 2019 | NCI Radiation Oncology Branch, NCI Center for Strategic Scientific Initiatives |
| 2nd Annual NRG Radiation Oncology Mini-Symposium: "AI and Machine Learning in Radiation Oncology" | January 10, 2020 | NRG Radiation Oncology Committee, NRG Center for Innovation in Radiation Oncology |
| NRG Oncology Digital Health Workshop | January 10, 2020 | NRG Oncology |

**Abbreviations**. AAPM: American Association of Physicists in Medicine; AI: Artificial Intelligence; ASTRO: American Society for Radiation Oncology; EORTC: European Organisation for Research and Treatment of Cancer, NCI: National Cancer Institute.

Radiation oncology-specific workshops (**Table 4**) such as the Practical Big Data Workshop[48] (hosted by the University of Michigan annually 2017-2019) and two ad hoc workshops by the NIH/NCI in 2019 (NCI Workshop on AI in Radiation Oncology in April[14] and the NCI State of Data Science in radiation Oncology in July[49]) and the have provided forums for data science practitioners to discuss results and ideas, but have not yet had a primary focus on education or research mentorship. A joint ASTRO/AAPM AI research workshop planned for June 2020 was cancelled due to the COVID-19 pandemic (https://www.astro.org/Meetings-and-Education/Live-Meetings/2020/Research-Workshop). Other promising avenues for future AI education and workshops include annual society (ASTRO, AAPM, etc.) and cooperative group meetings. For example, recent EORTC[50] and NRG Oncology meetings (https://www.nrgoncology.org/Resources/Meetings/January-2020-Semiannual-Meeting-Resources) have included workshops on AI/digital health and may be able to incorporate educational content. Another avenue to gain expertise during radiation oncology residency could be through the American Board of Radiology's B. Leonard Holman Research Pathway. This pathway is an established research fellowship during residency that provides between 18-21 protected months of research without lengthening clinical training time. This protected time could be used to gain expertise in data science, which could include MOOCs or AI fellowships in collaboration with data science departments or industry. Several companies, including Google, Microsoft, nVidia, and Facebook all offer 1-year AI "residencies" for specific areas such as deep learning. In late 2019, ASTRO and Varian Medical Systems announced a joint 1-year research fellowship to start in July 2020 for eligible residents (https://www.astro.org/Patient-Care-and-Research/Research/Funding-Opportunities/ASTRO-Varian-Award); research topics include AI, information systems, and related areas.

There are several grant opportunities for residents and fellows that could be utilized towards AI/ML research or education. These include 1-year grants of $25-50k by ASTRO (physicians and physicists), RSNA, and ASCO. There are also additional funding opportunities not specific to trainees by the Radiation Oncology Institute and radiotherapy companies. For new faculty, NIH K08/K23 awards can provide mentored research time and salary support. NIH R25 grants for developing informatics tools for cancer are another promising avenue for multi-year

funding opportunities. RSNA offers several grants for education (https://www.rsna.org/research/funding-opportunities/education-grants) and had a specific focus on developing AI education tools for their 2020 Education Innovation Grant.

## Conclusions

AI is becoming a transformative force in medicine but there are dangers to blindly trust trained models and raw data without understanding their governance. Just as radiation oncology trainees should understand requisite radiobiology and physics to treat patients, we believe that some level of competency in AI is necessary to safely and effectively utilize it in the clinical setting. In this perspective article from the 2019 NCI Workshop on AI in Radiation Oncology: Training and Education Working Group, we have discussed AI awareness and proper conduct (**Action Point 1**), what an AI curriculum might include (**AP2**), how to create and contribute to educational resources (**AP3**), and what support from institutions, societies, and funding agencies is required (**AP4**). We hope that this paper will spark further discussion on the future of trainee education in radiation oncology. One concrete path forward could be for radiation oncologists and medical physicists to collaboratively apply for fellowships/funding and develop workshops (**AP4**) for creation of data science educational curricula (**AP2**) and resources (**AP3**), while being mindful of the ethical concerns in AI implementation (**AP1**).

## Disclaimer